# Potential Structure and of a Plasma Hole in a Rotating Magnetized Plasma

Shinji Yoshimura[a], Atsushi Okamoto[b], and Masayoshi Y. Tanaka[a]

[a]National Institute for Fusion Science, Oroshi, Toki 509-5292, Japan
[b]University of Tokyo, 2-11-16 Yayoi, Bunkyo-ku, Tokyo 113-8656, Japan

## 1. Introduction

Electric field in magnetized plasmas drives a rotating motion by $E \times B$ drift, giving rise to macroscopic structures such as vortices. Structure formation in magnetized plasmas attracts many researchers, and a large number of theoretical and simulation studies have been performed. Recently spontaneous formation of a cylindrical density-cavity structure, which is referred to as *plasma hole*, has been observed in a laboratory magnetized plasma produced by electron cyclotron resonance heating. [1, 2] From the flow-structural point of view, the plasma hole is a monopole vortex with a sink and is identified as a dissipative vortex (Burgers vortex [3, 4]).

In this paper, another aspect of the plasma hole structure, i.e., the potential structure is presented. The maximum velocity of azimuthal $E \times B$ rotation is found to exceed the ion sound speed, which implies the existence of strong radial electric field. Here we present the results of potential profile measurement of the plasma hole. We evaluate the degree of quasi-neutrality breaking occurred in the plasma hole using Poisson's equation. Preliminary result of the potential fluctuation measurement is also presented.

## 2. Experimental

The experiments have been performed with the High Density Plasma Experiment (HYPER-I) device at National Institute for Fusion Science. [5] HYPER-I is a cylindrical plasma device (30 cm in diameter and 200 cm in length) with ten magnetic coils. The plasma is produced by electron cyclotron resonance heating in a *magnetic beach*, where the microwave power is 5 kW and the frequency is 2.45 GHz. A helium gas has been used with the operation pressure of 0.6 mmTorr. Typical plasma parameters are as follows: the electron temperature is 10–20 eV, the plasma density $10^{10}$cm$^{-3}$ (hole plasma) – $10^{11}$cm$^{-3}$ (ambient plasma).

Floating emissive probe method [6] has been adopted to measure the plasma potential profile. The emissive probe has been constructed as follows: the emissive filament is made of a small loop of tuntalum wire (0.23 mm in diameter) welded to tungsten rods (0.8 mm in diameter), which are mounted in a two-hole ceramic insulator. The measurement circuit is grounded through a high impedance load resister. As the electron emission from the heated filament becomes sufficiently high, the probe potential comes to indicate the plasma potential directly, which has been experimentally confirmed by the *I-V* characteristics of the emissive probe.

### 3. Results and Discussions

The perspective image taken by a CCD camera is shown in Fig. 1(a). The central dark region indicates the existence of a deep density hole, the sizes of which are 6 cm in diameter and more than 100 cm in axial length. This is why we call this structure *plasma hole*. The density in the hole plasma is indeed one tenth of that in the ambient plasma, and width of transition layer between the hole and ambient plasma is several ion Larmor radii (See Fig. 1(b)).

Two dimensional potential profile of the plasma hole is shown in Fig. 2 (a), in which two distinctive features can clearly be seen: (i) The potential has axisymmetric *bell-shaped* structure and sharply increases toward the center, the maximum value exceeding more than 80 V, which is about five times higher than the electron temperature. The drastic increase in plasma potential begins from the density transition layer (x ~ 30-50 mm), in which the density profile has the steepest gradient. On the other hand, the maximum value of the plasma potential without any characteristic structure is, as is expected, approximately equal to the electron temperature.

Since the plasma hole has the intense electric field (40 V/cm) compared to that without plasma hole (~1 V/cm), the breakdown of charge neutrality may takes place in the hole plasma. The quasi-neutrality breaking can be evaluated from the Poisson's equation $-\nabla^2 \phi = 4\pi e \delta n$, where $\delta n = n_i - n_e$. Assuming the magnitude of the potential $|\phi| \sim T_e / e$, and the characteristic scale-length is equal to plasma radius $L$, we have the normalized density difference as $\delta n / n \sim (\lambda_D / L)^2$, where $\lambda_D$ is the Debye length. A typical value of $\delta n / n$ for a plasma without plasma hole is estimated to be of the order of $10^{-6}$ under our experimental conditions ($T_e \sim 20$ eV, $L = 15$ cm and $n \sim 10^{11}$ cm$^{-3}$). It should be noted that $\delta n / n$ represents the degree of quasi-neutrality breaking, or non-neutrality, of the plasma. By taking the second derivative of measured potential profile, the value of $\delta n / n$ can be calculated directly. $\delta n / n$ of the plasma without plasma hole is indeed of the order of $10^{-6}$, which coincides with the expected value from the Poisson's equation. The degree of quasi-neutrality breaking of the plasma hole is, however, considerably greater than the expected value for a plasma without plasma hole (See Fig. 2(b)). The quantity $\delta n/n$ attains its maximum value ($5\times10^{-3}$) within the hole plasma region (ion-rich); it is more than $10^3$ times higher than that of quasi-neutral plasma.

Signals of the potential fluctuation measured at three different positions (r = 0 cm, 2.5 cm and 5.0cm) are shown in Fig. 3. Pronounced fluctuation is evident at r = 2.5 cm, the position of which corresponds to the interfacial layer between the hole and ambient plasma. Intermittent negative spikes are found in the hole region (r = 0 cm). Positive spikes in ion saturation current are simultaneously found in the hole region, which implies the existence of sporadic ion influx.

## 4. Conclusion

The characteristic potential profile of the plasma hole has been measured with an emissive probe. The potential has a bell-shaped structure and its maximum value is about five times higher than the electron temperature. The quantity $\delta n/n$ of the plasma hole has been calculated from the Poisson's equation and the potential data; it is about three orders of magnitude higher than that of the ambient plasma. It is found that the quasi-neutrality breaking occurs in the plasma hole, producing the very high potential ($\phi \sim 5T_e$) and the resultant supersonic ion flow in azimuthal motion. Preliminary results of potential fluctuation measurement reveal the pronounced fluctuation in the interfacial layer between the hole and ambient plasma. Moreover intermittent spiky signals are found in the hole region. Detailed study of potential fluctuations will form our future work.

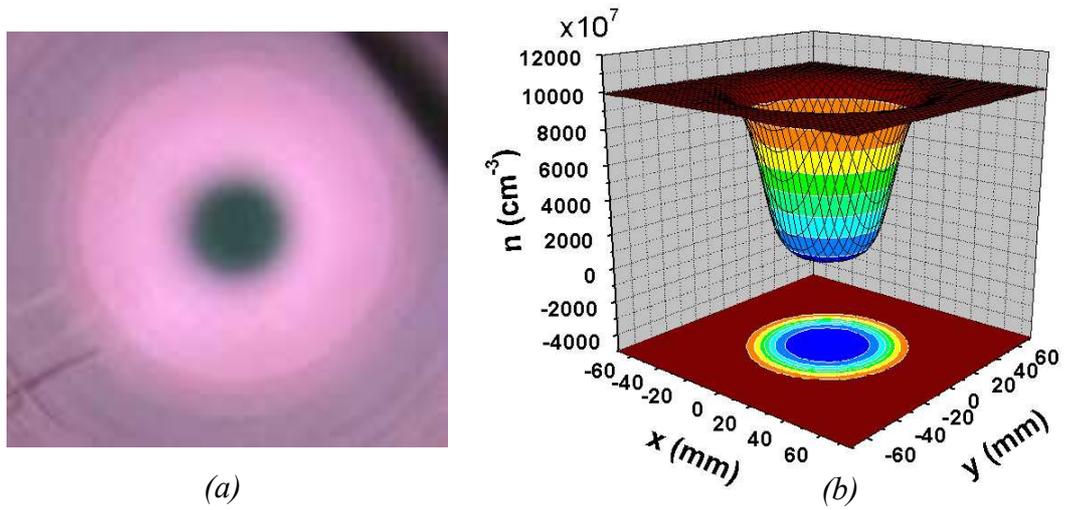

*Figure 1* (a) CCD image of the plasma hole. (b) Density profile of the plasma hole.

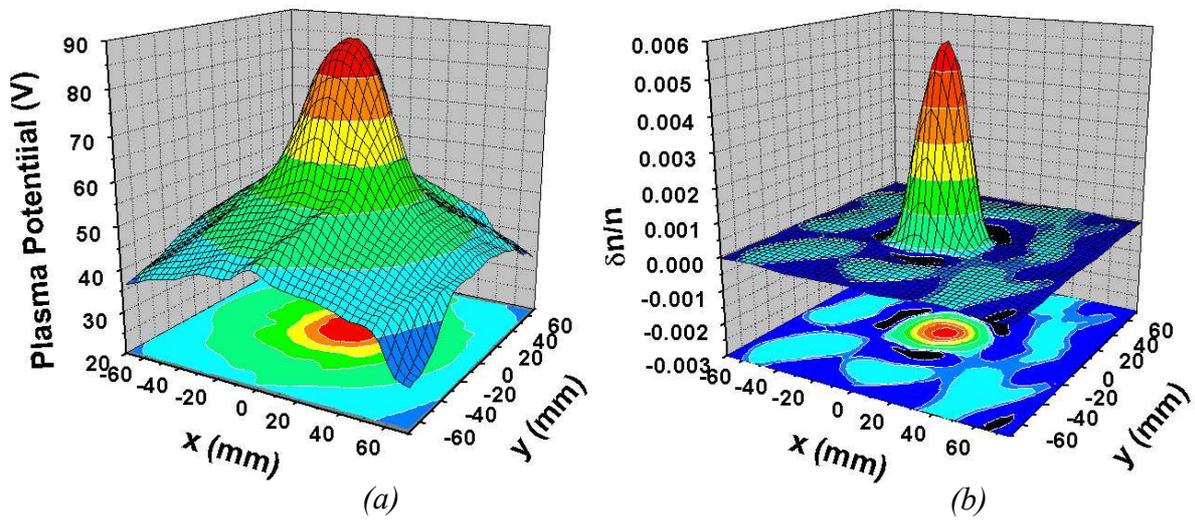

*Figure 2* (a) Potential profile of the plasma hole. (b) $\delta n / n$ profile of the plasma hole.

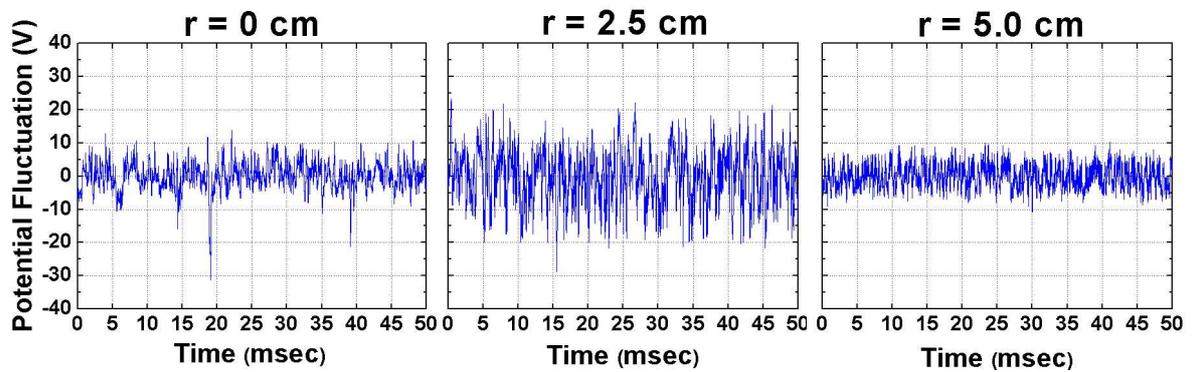

*Figure 3* Signals of potential fluctuation at r = 0 cm, 2.5 cm and 5.0 cm.